\documentclass[12pt]{article}
\usepackage{amssymb,amsmath}
\begin{document}


\sloppy
\title
{\Large  On the simplified tree graphs in gravity}
\author
 {
       A. I. Nikishov
          \thanks
             {E-mail: nikishov@lpi.ru}
  \\
               {\small \phantom{uuu}}
  \\
  \\
               {\it {\small} P.N.Lebedev Physical Institute, Moscow, Russia}
  \\
 }
%
\maketitle 
\begin{abstract}
Firstly, I give the reason why is wrong my previously made assumption that the volume integral over the pressure may not be zero in a system where the gravitation plays no role in holding the system together. Secondly, in the first nonlinear approximation I obtain  the inner and outer Schwarzschild solutions in harmonic and isotropic coordinates in two different ways. One way is to start from standard solution and make the appropriate coordinate transformation. The other way is to use the perturbation theory with elements of Schwinger and Weinberg source approach. This latter method is applicable in general case and it is useful to study all its peculiarities on known simple example such as Schwarzschild solution. It turns out that this method is simpler then S-metrics approach (previously made by Duff) and more informative as it it shows which contribution comes from what region of space. 
\end{abstract}
\section{introduction}
If the Earth gravitational field can be measured with $G^2-$ accuracy, than its changes, due to building up the pressure, leading to earthquakes, can be monitored. For this reason calculating the Earth gravitational field with this accuracy is of interest. Thus simplifying the perturbation methods is essential. As a first step in this direction we reproduce the corresponding approximation of the Schwarzschild solution by simple means.
  
In three-graviton vertex all gravitons enter symmetrically. It can be simplified 
in particular cases. For two on-shell gravitons this has been done in [1]. In 
our case one graviton is external field graviton. Using integration by parts we can 
release it from being differentiated. Then it interact with Weiberg gravitational 
energy-momentum tensor, see $\S6$ Ch. 7 in [2]. In static case this tensor gives (negative) gravitational energy density which is three times that of Newtonian one [3].
Yet, as shown in [3], for a pressure-free dust there is an over all agreement with Newtonian limit for
total energy of the system. We shall see later on that the same is true for the ball of liquid.

The negative energy density makes me uneasy. I prefer to think that this is only effective quantity. In Newtonian theory the gravitational energy density is given either by
$\frac12\mu\phi$  or by $ -\frac{(\nabla\phi)^2}{8\pi G}$, see $\S99$ in [4]. I think
that the correct expression should have the form $\mu\phi+\frac{(\nabla\phi)^2}{8\pi G}$ in which the gravitation field has positive energy density. The relativistic generalization of this
is given in [3].

In Section 2  I give the reason why is wrong my previously made assumption that the volume integral over the pressure may not be zero in a system where the gravitation plays no role in holding the system together.

In Section 3 the interior Schwarzschild solutions in harmonic and isotropic coordinates (in $G^2-$approximation) are obtained from the standard solution. They have the forms different from those obtained by using tree graphs or solving 
Einstein equations by perturbation procedure. Yet these differences are only in forms.

In Section 4 the Weiberg gravitational energy-momentum tensor and relevant expressions
for uniform liquid of a ball are given.

In Section 5 the nonlinear corrections are obtained.

\section{ Gravitational field of a system in which gravitation plays no role}

For a statical system the metric at large distance (in comparison with its size) is
determined by the integral over energy-momentum tensor. In particular we need the integral $\int \sigma_{ik}dV$ where $\sigma_{ik}$ is the stress tensor.This integral 
is equal to $\int P_i x_k df$ where $P_i$ is the force acting on the surfice of the body and the integral is over the surfice, see $\S3$ in [5]. As in our case $P_i=0$, the integral is zero and pressure plays no role in generating the gravitational field at large distances This assertion correct the mistaken assumption in [6].

\section{Interior Schwarzschild solutions in harmonic and isotropic coordinates}
Here we obtain the solution in harmonic coordinates from the known solution in
standard coordinates. The metric have the form, see Synge [7], eq.(7.183)
$$
ds^2=\frac{dr^2}{1-qr^2}+r^2(d\theta^2+\sin^2\theta d\varphi^2)-
\{\frac{3\sqrt{1-qa^2}-\sqrt{1-qr^2}}{2}\}^2dt^2,\quad q=\frac{2mG}{a^3}. \eqno(1)
$$
In this Section we denote the harmonic coordinate radius by $R$ and use the substitution
$$
r=R(1-\phi(m.b.R)),\quad \phi(m.b.R)=\frac{mG}{2b}(-3+\frac{R^2}{b^2}). \eqno(2)
$$
The radius of the liquid ball $a$ in standard coordinates is related to that in harmonic coordinates $b$ by the expression
$$
a=b+mG.                                                                    \eqno(3)
$$
We are working in $G^2-$approximation. It is easy to obtain
$$
\frac{dr^2}{1-qr^2}=[1+\frac{mG}{b}(3-\frac{R^2}{b^2})+\frac{m^2G^2}{b^2}(\frac{4}{9}+
\frac{3}{2}\frac{R^2}{b^2}-\frac74\frac{R^4}{b^4})]dR^2+O(G^3).           \eqno(4)
$$
In terms $\propto G^2$ we have replaced $a$ by $b$.
Using the relations
$$
RdR=X_{\alpha}dX_{\alpha}, \quad (dR)^2=\frac{X_{\alpha}X_{\beta}dX_{\alpha}dX_{\beta}}{R^2}, \quad \alpha,\beta=1,2,3.,                                                                            \eqno(5)
$$
we find, see (2) and (3)
$$
g_{\alpha\beta}(\vec X)=[1-\phi(m,b,R)]^2\delta_{\alpha\beta}+\frac{m^2G^2}{b^2}
(\frac{3R^2}{b^2}-\frac{2R^4}{b^4})\frac{X_{\alpha}X_{\beta}}{R^2}.       \eqno(6)
$$
For $g_{00}$ we obtain from (1)
$$
g_{00}=-[1+2\phi(m,a,r)+\frac{m^2G^2}{b^2}(\frac34-\frac32\frac{R^2}{b^2}+\frac34\frac{R^4}{b^4})].                                                             \eqno(7)
$$
Here again we have replaced $a$ by $b$
and $r$ by $R$ in terms $\propto G^2$. Expressing $a$ and $r$ in terms of $b$
and $R$ we find
$$
2\phi(m,a,r)\equiv \frac{mG}{a}(-3+\frac{r^2}{a^2})= 2\phi(m,b,R)+\frac{3m^2G^2}{b^2}-\frac{m^2G^2R^4}{b^6}.                     \eqno(8)
$$
So, from (7) and (8) we have
$$
g_{00}=-\{1+2\phi(m,b,R)+\frac{m^2G^2}{b^2}(\frac{15}{4}-\frac32\frac{R^2}{b^2}-
\frac{R^4}{4b^4})\}.                                                          \eqno(9)
$$

In isotropic coordinate system in the considered approximation $b$ remains the same.
Indeed, if we denote (only in the formula below) the radius of the ball in isotropic system as $b$,
then
$$
a=b(1+\frac{mG}{2b})^{2}\approx b+mG,                                      \eqno(10)
$$
$g_{00}$ in isotropic coordinates has the same form (9) and $g_{\alpha\beta}$ can be obtained from (6) by
adding pure gauge functions with appropriate coefficients. The gauge function has the 
form 
$$
(\Lambda_{\alpha})_{,\beta}+(\Lambda_{\beta})_{,\alpha}.                 
$$

{\sl In what follows we replace $R$ by $r$}. We need three such functions
$$
(x_{\alpha})_{\beta}=\delta_{\alpha\beta},\quad (x_{\alpha}r^2)_{,\beta}=
\delta_{\alpha\beta}r^2+2x_{\alpha}x_{\beta},\quad (x_{\alpha}r^4)_{,\beta}=
\delta_{\alpha\beta}r^4+4r^2x_{\alpha}x_{\beta},                            \eqno(11)
$$
Adding to r.h.s. of (6) the function
$$
h^{gauge}_{\alpha\beta}=\frac{m^2G^2}{b^2}\{\frac64\delta_{\alpha\beta}-
\frac{3}{2b^2}(\delta_{\alpha\beta}r^2+2x_{\alpha}x_{\beta})+
\frac{1}{2b^4}(\delta_{\alpha\beta}r^4 +4r^2x_{\alpha}x_{\beta})\},        \eqno(12)
$$
 we get in isotropic coordinate system
$$
g_{\alpha\beta}(m,r)=\delta_{\alpha\beta}[1+\frac{mG}{b}(3-\frac{r^2}{b^2})+
\frac{m^2G^2}{b^2}(\frac{15}{4}-\frac{3r^2}{b^2}+\frac{3r^4}{4b^4})],\quad r<b.      \eqno(13)
$$
In both harmonic and isotropic systems the metric and their first derivatives are continuous
across the boundary of the ball, see also [9] and [11].

\section{Weinberg gravitational energy-momentum tensor}

In general relativity the nonlinear source of metric is the Weinberg gravitational energy-momentum tensor, see $\S7$ in [2]
$$
\tau_{ik}=T_{ik}+t_{ik},\quad i,k=0,1,2,3,                                                 
$$
$$
t_{ik}=\frac{1}{8\pi G}[-\frac12h_{ik}R^{(1)}+\frac12\eta_{ik}h^{lm}R^{(1)}_{lm}+
R^{(2)}_{ik}-\frac12\eta_{ik}R^{(2)}].                                  \eqno(14)
$$
Here $T_{ik}$ is energy -momentum tensor of matter, $R^{(1)}$ and
$R^{(2)}$ are linear and quadratic in $h$ parts of $R$,
$$
g_{ik}=\eta_{ik}+h_{ik},\quad h_{ik}=\stackrel{1}{h}_{ik}+\stackrel{2}h_{ik}+O(G^3)                                                                                                                                                                    \eqno(15) 
$$
$$
 \eta_{ik}=\rm diag(-1,1,1,1) \quad
R^{(1)}=R^{(1)}_i{}^i,\quad R^{(2)}=R^{(2)}_i{}^i=\eta^{ik}R^{(2)}_{ik},
$$
$$
R^{(1)}_{ik}=\frac12(h^j{}_{j,ik}-h^j{}_{i,jk}-h^j{}_{k,ji}+h_{ik,j}{}^j  ). \eqno(16)
$$
The expression for $\stackrel{2}{R}{}^{(2)}_{ik}$ see eq.(7.6.15) in [2] (or in another form in [3]).

Using the expression for $t_{\alpha\beta}$ in (33) below, it is easy to verify that
$$
t^{\alpha\beta}{}_{,\beta}=\frac{1}{4\pi G}\phi^{,\alpha}\phi_{,\beta\beta}=\mu\phi^{,\alpha},\quad \alpha, \beta=1, 2, 3.                     \eqno(17)
$$
Hence 
$$
(T^{\alpha\beta}+t^{\alpha\beta})_{\beta}=p^{,\alpha}+\mu\phi^{,\alpha}=0.  \eqno(18)
$$
This is the equation for equilibrium of a liquid ball, see  \S 3 in [8].
For $\mu=const$ differentiation over $x^{\alpha}$
gives $\nabla^2p=-4\pi G\mu^2$ or  in spherical coordinates
$$
\frac{1}{r^2}\frac{d}{dr}(r^2\frac{dp}{dr})=-4\pi G\mu^2.                 \eqno(19)
$$
Twice integrating over $r$ from $0$ to $r$ we find
$$
p(r)-p(0)=-\frac132\pi G\mu^2r^2.                                            \eqno(20)
$$
As $p(b)=0$ we have from here
$$
p(0)=\frac23\pi G\mu^2b^2,                                               \eqno(21)
$$
so that 
$$
p(r)=\frac23\pi G\mu^2(b^2-r^2)=\frac{1}{8\pi G}\frac{m^2G^2}{b^4}(3-\frac{3r^2}{b^2}).
                                                                          \eqno(22)
$$
Here we neglect the difference between $b$ and $a$ and difference between $m_0$ and $m$,
see below.

The Einstein equations in this formalism have the form [2]
$$
R^{(1)}_{ik}-\frac12\eta_{ik}R^{(1)}=-8\pi G(T_{ik}+t_{ik}).                 \eqno(23).
$$

 To calculate $t_{ik}$ we need $h^{(1)}_{ik}$. It is the same in harmonic and isotropic coordinates:
 $$
 \stackrel{1}{h}_{ik}=-2\phi\delta_{ik,},\quad \stackrel{1}h\equiv \stackrel{1}{h}_i{}^i=-4\phi.                                            \eqno(24)
 $$
Here $\phi$ is the Newtonian potential, see (26) below.

For the ball of liquid with uniform density we have
$$
T_i{}^j=(\mu+p)u_iu^j+p\delta_i{}^j,\quad T_0{}^0=-\mu\theta(b-r).      \eqno(25)
$$
Here $b$ is the radius of the ball in the considered coordinate system.
Now we have 
  $$
    \phi(m',r)=\frac{m'G}{2b}(-3+\frac{r^2}{b^2}),\quad r<b,           \eqno(26a)
   $$
$$
    \phi(m',r)=-\frac{m'G}{r}.\quad r>b.                                    \eqno(26b)
$$
Here
$$
m'=\frac43\pi\mu b^3.                                                       \eqno(27)
$$

We remind here that in general relativity the dressed mass is given by the expression [8], [9]
$$
m=\frac43\pi\mu a^3,                                                       \eqno(28) 
$$
where $a$ is the ball radius in standard Schwarzschild
coordinates.The relation $a=b+mG$ is exact in harmonic coordinates and approximate in isotropic one. The bare mass is, see \S 100 in [7], or [9]
$$
  m_0 =\int dV\mu e^{\lambda(r)/2}=\int dV\mu(1+\frac43\pi\mu r^2G)=   =m+\frac{3m_0^2G}{5b}.                                                 \eqno(29)
$$
From (27), (28) and (3) we have [9]
$$
m'=m(1-\frac{3mG}{b})=m_0(1-\frac{18m_0G}{5b.}).                        \eqno(30)
$$
Now, in terms of $\phi$ (see (26)) tensor $t_{ik}$ in (14) has the form
 $$
 t_{ik}=\frac{1}{8\pi G}\{\eta_{ik}[3(\nabla \phi)^2+4\phi\phi_{,\alpha\alpha }]-
  2\phi_{,i}\phi_{,k}-4\phi\phi_{,ik}-8\phi\phi_{,\alpha\alpha}\delta_{i0}\delta_{k0}\},\eqno(31)
$$  
$$
 \phi_{,i}=\frac{\partial \phi}{\partial x^i},\quad     \phi_{,\alpha\alpha}=\nabla^2\phi=4\pi G\mu.                          
$$
The indices of (true in the sense of general relativity) tensor $T_i{}^k$ are raised and lowed with $g$ not with $\eta$ [2]. So, from (19) we have
$$
T_0{}^0=-\mu,\quad T_{\alpha}{}^{\beta}=\delta_{\alpha\beta}p,\quad 
g_{00}T_0{}^0=\mu+2\mu\phi=T_{00}^{(0)}+T_{00}^{(1)},\quad
 T_{00}^{(0)}=\mu,\quad  T_{00}^{(1)}=2\mu\phi,                \eqno(32)
$$
and from (25)
$$
t_{00}=-\frac{3}{8\pi G}(\nabla\phi)^2-6\mu\phi,
$$
$$
t_{\alpha\beta}=\frac{1}{8\pi G}\{
\delta_{\alpha\beta}[3(\nabla\phi)^2+4\phi\phi_{,\gamma\gamma}] -2\phi_{,\alpha}\phi_{,\beta}-4\phi\phi_{,\alpha\beta}   \}.                                                                                                   \eqno(33)
$$

 We also need the quantities
$$
\bar t_{00}=t_{00}-\frac12\eta_{00}t=\frac{1}{4\pi G}(\nabla\phi)^2-\mu\phi,
$$
$$
t=t_{\alpha\alpha}-t_{00}=\frac{5}{4\pi G}(\nabla\phi)^2+10\mu\phi.    \eqno(34)
$$
$$
\bar T_{00}=T_{00}-\frac12g_{00}T=\frac12\mu+\frac32p+\mu\phi\equiv\bar T_{00}^{(0)}+
\stackrel{1}{\bar T}{}_{00}.                                                                                                                                       \eqno(35)
$$

 Using (26) for $r<b$ we get
$$
t=\frac{1}{4\pi G}\frac{m^2G^2}{b^4}(-45+\frac{20r^2}{b^2}),\quad
t_{00}=\frac{1}{8\pi G}\frac{m^2G^2}{b^4}(54-\frac{21r^2}{b^2}),
$$
$$
  \bar t_{00}=\frac{1}{8\pi G}\frac{m^2G^2}{b^4}(9-\frac{r^2}{b^2}),       \eqno(36)
$$.       
$$
t_{\alpha\beta}=\frac{1}{8\pi G}\frac{m^2G^2}{b^4}\{\delta_{\alpha\beta}(-12
+\frac{7r^2}{b^2})-
\frac{2x_{\alpha}x_{\beta}}{b^2}\},                                         \eqno(37)
$$
$$
T_{00}^{(1)}=2\mu\phi=\frac{1}{4\pi G}\frac{m^2G^2}{b^4}(-9+\frac{3r^2}{b^2}).                                                                               \eqno(38)
$$
$$
\stackrel{1}{\bar T}{}_{00}=\frac32p+\mu\phi=\frac{1}{16\pi G}\frac{m^2G^2}{b^4}(-9-\frac{3r^2}{b^2}),                         \eqno(39)
$$
Similarly, for $r>b$ we have
$$
t_{00}=-\frac{3}{8\pi G}\frac{m^2G^2}{r^4}, \quad \bar t_{00}=\frac{1}{4\pi G}\frac{m^2G^2}{r^4}, \quad t=\frac{5}{4\pi G}\frac{m^2 G^2}{r^4},   \eqno (40)
$$
$$
t_{\alpha\beta}=\frac{1}{8\pi G}\frac{m^2G^2}{r^4}\{7\delta_{\alpha\beta}-
                             \frac{14x_{\alpha}x_{\beta}}{r^2}\},      \eqno(41)       
 $$
Knowing $t_{00}$  (see (36) and (40))and $T_{00}^{(1)}$  (see(38)) we can calculate the total gravitational energy $U$:
$$
U= \int dV(t_{00}+T_{00}^{(1)})=(\frac{69}{10}  -\frac32-\frac{12}{5}-\frac{18}{5})\frac{m^2G}{b}=-
\frac35\frac{m^2G}{b}.                                             \eqno(42)
$$ 
The first number (i.e. $\frac{69}{10}$) comes from $t_{00}$ when $r<b$, the second
($-\frac32$) when $r>b$, the third term comes from $T_{00}
^{(1)}$ and the last one from $m'$, see eq.(30). In the Newtonian theory we have
$$
U=-(\frac12+\frac{1}{10})\frac{m^2G}{b}=-\frac35\frac{m^2G}{b}..                                                                                                                                                                                        \eqno(43)
$$  
The number $\frac12$ comes from $,r>b$ and $\frac{1}{10}$ from $r<b$.
In terms $\propto G^2$ we neglect the difference between $m$ and $m_0$.
  
\section{ Nonlinear corrections to metric}

Up to gauge terms $\stackrel{2}{h}_{ik}(m',r)$ can be obtained from
the formula (cf. $\S17$ in [10])
$$
\stackrel{2}{h}_{ik}(m',x)=\int d^4x'D_{+}(x-x')\theta(x'),\quad \eqno(44)
$$
$$
 \theta_{ik}(x')=16\pi G(\stackrel{1}{\bar T}_{ik}(x')+\bar t_{ik}(x')), \quad \eqno(45)                                                                     
$$
Here
$$
D_{+}(x)=\frac{1}{4\pi}\delta_+(x^2)=\frac{1}{(2\pi)^2}\frac{1}{x^2+i\epsilon},\quad
\partial_i\partial^iD_+(x)=-\delta(x).                                      \eqno(46)
$$
From (44), (46) we have
$$
\partial_j\partial^j \stackrel{2}{h}_{ik}(m',r)=-\theta_{ik}(x),   \eqno(47)
$$
or in our case when $\stackrel{2}{h}_{ik}(m',r)$ is independent of time
$$
\nabla^2\stackrel{2}{h}_{ik}(m',r)=-\theta_{ik}(x),   \eqno(48)
$$
We note also that 
$$
\int d\tau D_+(\vec x-\vec x',\tau)=\frac{1}{4\pi|\vec x-\vec x'|},   \eqno(49)
$$
see eq.(3.92), Ch.2 in [10].

Using also the relation
$$
\int_{-1}^1\frac{dt}{\sqrt{r^2-r'^2-2rr't}}=
\begin{cases}
\frac2r, & \text{if }r'<r\\
\frac{2}{r'}, & \text{if }r'>r,
\end{cases}                                                   \eqno(50)
$$
it is easy to find that for any function $f(r)$ we have
$$
\int_{r'>c}d^4x'D_+(x-x')f(r')=\frac1r\int_c^rdr'r'^2f(r')+\int_r^{\infty}dr'r'f(r').
                                               \eqno(51)
$$

Now for $r<b$, adding $\bar t_{00}$ in (36) and $\stackrel{1}{\bar T}{}_{00}$ in (39) we have
$$
\frac{1}{16\pi G}\theta_{00}=\bar t_{00}+\stackrel{1}{\bar T}{}_{00}=\frac{1}{16\pi G}\frac{m^2G^2}{b^4}(9-\frac{5r^2}{b^2}).                   \eqno(52)
$$
Using also relations
$$
\int d^4x'D_+(x-x')\theta(b-r')=\begin{cases}
\frac12b^2-\frac16r^2, & \text{if }r<b\\
\frac{b^3}{3r}, & \text{if }r>b,
\end{cases}                                                   \eqno(53)
$$
$$
\int d^4x'D_+(x-x')r'^2\theta(b-r')=\begin{cases}
\frac14b^4-\frac{1}{20}r^4, & \text{if }r<b\\
\frac{b^5}{5r}, & \text{if }r>b,
\end{cases}                                                   \eqno(54)
$$
we find the contribution to $\stackrel{2}h_{00}$ from $r'<b$
$$
\int d^4x'D_+(x-x')\frac{m^2G^2}{b^4}(9-\frac{5r'^2}{b^2})\theta(b-r')=
\frac{m'{}^2G^2}{b^2}(\frac{13}{4}-\frac32\frac{r^2}{b^2}+
\frac{r^4}{b^4}), \quad   r<b.                                           \eqno(55)
$$
We note that r.h. sides of (53) and (54) are continuous at $r=b$ 

For $r'>b$ we have $\bar T_{00}^{(1)}=0$ and $\bar t_{00}$ is given in (40).
So the contribution to $\stackrel{2}h_{00}$ is
$$
16\pi G\int_{r'>b}d^4x'D_+(x-x')\frac{m^2G^2}{4\pi Gr'^4}=\frac{2m^2G^2}{b^2},\quad 
r<b.                                                                    \eqno(56)
$$
Addind (55) and (56), we find 
$$
\stackrel{2}{h}_{00}(m',r)=\frac{m{}^2G^2}{b^2}(\frac{21}{4}-\frac32\frac{r^2}{b^2}+
\frac{r^4}{b^4}), \quad   r<b.                                             \eqno(57)     $$

Next, we express $h_{00}^{(1)}(m',r)$ in terms of $m$, see eqs. (24), (26) and (30)
$$
\stackrel{1}{h}_{00}(m',r)=-2\phi(m',r)=\stackrel{1}{h}_{00}(m,r)+\frac{m^2G^2}{b^2}(
-9+\frac{r^2}{b^2}),
\quad r<b.                                                            \eqno(58)
$$
Thus,
$$
\stackrel{1}{h}_{00}(m',r)+\stackrel{2}{h}_{00}(m',r)=\stackrel{1}{h}_{00}(m,r)+\stackrel{2}{h}_{00}(m,r),                                                       \eqno(59)
$$
where in agreement with (9)
$$
\stackrel{2}{h}_{00}(m,r)=\frac{m^2G^2}{b^2}(-\frac{15}{4}+\frac{3r^2}{2b^2}+
\frac{r^4}{4b^4}), \quad  r<b.                                                                                                                                   \eqno(60)
$$
For $r>b$ the contribution to $\stackrel{2}{h}_{00}$  from $r'>b$ is
$$
16\pi G\int_{r'>b}d^4x'D_+(x-x')\frac{m^2G^2}{4\pi Gr'^4}=m^2G^2(\frac{4}{rb}-\frac{2}{r^2}), \quad  r>b.                     \eqno(61)
$$
The contribution from $r'<b$ is
$$
\int d^4x'D_+(x-x')\frac{m^2G^2}{b^4}(9-\frac{5r^2}{b^2})\theta(b-r')=
m^2G^2\frac {2}{rb}, \quad r>b.                            \eqno(62)
$$
Thus, the sum of these two contributions is
$$
\stackrel{2}{h}_{00}(m',r)=m^2G^2(\frac{6}{rb}-\frac{2}{r^2}), \quad  r>b. \eqno(63)
$$

On the other hand, from (24), (26) and (30) we have
$$
\stackrel{1}{h}_{00}(m',r)=-2\phi(m',r)=\frac{2m'G}{r}=\frac{2mG}{r}(1-\frac{3mG}{b}), \quad r>b                                                              \eqno(64)
$$
Adding (63 and (64), we see that eq.(59) holds also for $r>b$.

From (59) and (60), we have 
$$
g_{00}(m,r)=-1-2\phi(m,r)+\frac{m^3G^2}{b^2}(-\frac{15}{4}+\frac{3r^2}{2b^2}+
\frac{r^4}{4b^4}), \quad r<b.                                       \eqno(65)
$$
This agrees with Rosen [11], when terms up to $G^2$ are retained in his eq.(15).

For $r>b$, we find
$$
g_{00}(m,r)=-1+\frac{2mG}{r}-\frac{2m^2G^2}{r^2}, \quad  r>b.             \eqno(66)
$$

To calculate $\stackrel{2}{h}_{\alpha\beta}(m',r)$ for $r<b$ we note that it should 
have the form
$$
\stackrel{2}{h}_{\alpha\beta}(m',r)=\delta_{\alpha\beta}\frac{m^2G^2}{b^2}(c_0+c_2\frac{r^2}{b^2}+c_4\frac{r^4}{b^4}).
.                                                                         \eqno(67)
$$
From here with the help of (57) we find
$$
\stackrel{2}{h}(m',r)=\stackrel{2}{h}_{\alpha\alpha}(m',r)-\stackrel{2}{h}_{00}(m',r)
=\frac{m^2G^2}{b^2}\{(3c_0-\frac{21}{4})+(3c_2+\frac32)\frac{r^2}{b^2}+(3c_4-\frac14)\frac{r^4}{b^4}\},                                                           \eqno(68)
$$
and using the relation 
$$(r^n)_{,\alpha\alpha}=(n^2+n)r^{n-2}                                   \eqno(69)
$$
 for $n=2$, and $n=4$ we get from (68)
$$
\nabla^2\stackrel{2}{h}(m',r)
=\frac{m^2G^2}{b^2}\{(3c_2+\frac32)\frac{6}{b^2}+(3c_4-\frac14)\frac{20r^2}{b^4}\}.                                                      \eqno(70)
$$

From (67) we obtain
$$
( \stackrel{2}{h}_{\alpha\beta}(m',r))_{,\alpha\beta}=\frac{m^2G^2}{b^4}(6c_2+
20c_4\frac{r^2}{B^2}).                                                  \eqno(71)
$$
Thus,
$$
\stackrel{2}{R}{}^{(1)}(m',r)=\nabla^2\stackrel{2}{h}(m',r)-(\stackrel{2}{h}_{\alpha\beta}(m',r))_{,\alpha\beta}=\frac{m^2G^2}{b^4}\{(12c_2+9)+(40c_4-5)\frac{r^2}{b^2}\}.                                                                              \eqno(72)
$$

On the other hand, we  can obtain $\stackrel{2}{R}{}^{(1)}$ from the Einstein equation
$$
R=8\pi GT_i{}^i=8\pi G(-\mu+3p).                                        \eqno(73)
$$
In our case 
$$
R=\stackrel{1}{R}{}^{(1)}+\stackrel{2}{R}{}^{(1)}+\stackrel{2}{R}{}^{(2)}-
h^{ik}\stackrel{1}{R}{}^{(1)}_{ik}, \quad \stackrel{1}{R}{}^{(1)}=-2\phi_{,\alpha\alpha}
=-8\pi G\mu.                                                            \eqno(74)
$$
So, 
$$
\stackrel{2}{R}{}^{(1)}=8\pi G3p-\stackrel{2}{R}{}^{(2)}+h^{ik}\stackrel{}{R}{}^{(1)}_{ik},\quad h^{ik}\stackrel{}{R}{}^{(1)}_{ik}=8\phi\phi_{,\alpha\alpha}=\frac{m^2G^2}{b^4}
(-36+\frac{12r^2}{b^2}),                                               \eqno(75)
$$
$$
\stackrel{2}{R}{}^{(2)}=-8\phi\phi_{,\alpha\alpha}-10(\nabla\phi)^2=\frac{m^2G^2}{b^4}
(36-\frac{22r^2}{b^2}),.                                               \eqno(76)                                     
$$
Collecting all these terms, we obtain
$$
\stackrel{2}{R}{}^{(1)}=\frac{m^2G^2}{b^4}
(-63+\frac{25r^2}{b^2}),                                                   \eqno(77)  $$
Comparison with (72) gives $c_2=-6$, $c_4=3/4$.

Now, using (67) and (69), with these $c_2, c_4$ we can find the source of 
$\stackrel{2}{h}_{\alpha\beta}(m',r)$:
$$
\nabla^2\stackrel{2}{h}_{\alpha\beta}(m',r)=
\delta_{\alpha\beta}\frac{m^2G^2}{b^4}(-36+\frac{15r^2}{b^2}).                \eqno(78)
$$
For $r<b$ this source contributes to $\stackrel{2}{h}_{\alpha\beta}(m',r)$:
$$
\int d^4x'D_+(x-x')\delta_{\alpha\beta}\frac{m^2G^2}{b^4}(36-\frac{15r'^2}{b^2})
\theta(b-r')=\delta_{\alpha\beta}\frac{m^2G^2}{b^2}(\frac{57}{4}-\frac{6r^2}{b^2}+\frac{3r^4}{4b^4}).                                                                                                           \eqno(79)
$$
For $r>b$ its contribution is, see (53) and (54)
$$
\int d^4x'D_+(x-x')\delta_{\alpha\beta}\frac{m^2G^2}{b^4}(36-\frac{15r'^2}{b^2})\theta(b-r')
=9\delta_{\alpha\beta}\frac{m^2G^2}{br}, \quad r>b                                                                                                                  \eqno(80)
$$

For $r>b$ we have (see also [3])
$$
\bar{t}_{\alpha\beta}=t_{\alpha\beta}-\frac12\eta_{\alpha\beta}t=
\frac{1}{16\pi G}m^2G^2(\frac{4\delta_{\alpha\beta}}{r^4}-\frac{28x_{\alpha}x_{\beta}}{r^4}).
 \quad r>b.                                                              \eqno(81)
$$
But this source generates the function, see [3]
$$
\stackrel{2}{h}_{\alpha\beta}=m^2G^2(\frac{5\delta_{\alpha\beta}}{r^2}-
\frac{7x_{\alpha}x_{\beta}}{r^4}),                                         \eqno(82)
$$
which differs from the isotropic one by a multiple of the gauge function
$$
\Lambda_{\alpha,\beta}=m^2G^2(\frac{\delta_{\alpha\beta}}{r^2}-
\frac{2x_{\alpha}x_{\beta}}{r^4}).                                       \eqno(83)
$$
From the relation
$$
\nabla^2\Lambda_{\alpha,\beta}=m^2G^2(-2\frac{\delta_{\alpha\beta}}{r^4}+
\frac{8x_{\alpha}x_{\beta}}{r^6})                                       \eqno(84)
$$
we see that adding to $\bar{t}_{\alpha\beta}$ in (81) the gauge source
$$
\frac72\frac{1}{16\pi G}\nabla^2\Lambda_{\alpha,\beta}=\frac{1}{16\pi G}m^2G^2(-7\frac{\delta_{\alpha\beta}}{r^4}+
\frac{28x_{\alpha}x_{\beta}}{r^6})                                       \eqno(85)
$$
gives the source of $\stackrel{2}{h}_{\alpha\beta}(m',r)$ in isotropic coordinates:
$$
\theta_{\alpha\beta}=-3m^2G^2\frac{\delta_{\alpha\beta}}{r^4},\quad r>b.       \eqno(86)
$$
Thus , for $r'>b$ and $r<b$ the contribution to $\stackrel{2}{h}_{\alpha\beta}$ from this source is
$$
\int_{r'>b} d^4x'D_+(x-x')m^2G^2\frac{-3\delta_{\alpha\beta}}{r'^4}=-3\delta_{\alpha\beta}
m^2G^2\int_b^{\infty}dr'\frac{r'}{r'^4}=-\frac32\delta_{\alpha\beta}\frac{m^2G^2}{b^2}, \quad r<b.                                                                          \eqno(87)
$$
Finally, adding this to (79) gives 
$$
\stackrel{2}{h}_{\alpha\beta}(m',r)=\delta_{\alpha\beta}\frac{m^2G^2}{b^2}(\frac{51}{4}-\frac{6r^2}{b^2}+\frac{3r^4}{4b^4}),\quad  r<b.                                                                                                                  \eqno(88)
$$

Using (24), (26a) and first eq. in (30) we express $\stackrel{1}{h}_{\alpha\beta}(m',r)$ in terms 
of $m$:
$$
\stackrel{1}{h}_{\alpha\beta}(m',r)=\stackrel{1}{h}_{\alpha\beta}(m,r)+
\delta_{\alpha\beta}\frac{m^2G^2}{b^2}(-9+\frac{3r^2}{b^2}),\quad r<b.  \eqno(89)
$$
From this and (88) we have
$$
\stackrel{1}{h}_{\alpha\beta}(m',r)+\stackrel{2}{h}_{\alpha\beta}(m',r)=
\stackrel{1}{h}_{\alpha\beta}(m,r)+\stackrel{2}{h}_{\alpha\beta}(m,r),\quad r<b.
                                                                         \eqno(90)
$$
Here
$$
\stackrel{2}{h}_{\alpha\beta}(m,r)=\delta_{\alpha\beta}\frac{m^2G^2}{b^2}(\frac{15}{4}-\frac{3r^2}{b^2}+\frac{3r^4}{4b^4}),\quad  r<b.                                                                                                                  \eqno(91)
$$

For $r>b$ the contribution from the source (86) is, see (51)
$$
\int_{r'>b} d^4x'D_+(x-x')m^2G^2\frac{-3\delta_{\alpha\beta}}{r'^4}=-3\delta_{\alpha\beta}
m^2G^2\{\frac1r\int_b^rdr'\frac{r'^2}{r'^4}+\int_r^{\infty}dr'\frac{r'}{r'^4}\}=
$$
$$
-3\delta_{\alpha\beta}
m^2G^2\{\frac{1}{rb} -\frac{1}{2r^2}\}, \quad r>b.                                                                                                           \eqno(92)
$$
Adding  (80) and (92) we find
$$
\stackrel{2}{h}_{\alpha\beta}(m',r)=\delta_{\alpha\beta}
m^2G^2\{\frac{6}{rb} +\frac{3}{2r^2}\}, \quad r>b.                       \eqno(93)
$$

Using the first eq. in (30), we may rewrite  $\stackrel{1}{h}_{\alpha\beta}(m',r)$ in (24), (26b) as follows
$$
\stackrel{1}{h}_{\alpha\beta}(m',r)=\delta_{\alpha\beta}\frac{2m'G}{r}=
\delta_{\alpha\beta}(\frac{2mG}{r}-6m^2G^2\frac{1}{rb}),\quad r>b.                                                                                          \eqno(94)
$$
From the sum of the last two equations we have
$$
\stackrel{1}{h}_{\alpha\beta}(m',r)+\stackrel{2}{h}_{\alpha\beta}(m',r)=
\stackrel{1}{h}_{\alpha\beta}(m,r)+\stackrel{2}{h}_{\alpha\beta}(m',r).\quad r>b.
                                                                      \eqno(95)
$$
Here
$$
\stackrel{2}{h}_{\alpha\beta}(m,r)=\delta_{\alpha\beta}\frac{3m^2G^2}{2r^2},\quad r>b.
                                                                              \eqno(96).$$

\section{Concluding remarks}

One specific feature of the considered approach is that the metrics at any point is formed by the nonlinear sources in all space. In standard approach the inner and outer Schwarzschild solutions are obtained separately in each region.

The second feature is the appearance of mass $m'$ which depends on coordinate system.
It can be expressed via invariant ``dressed`` mass $m$ and then the solutions take 
a more simple form [9].

The gravitational energy density $t_{00}$ of the Weinberg tensor $t_{ik}$ is not the Newtonian one. This may indicate that in the future the gravitational theory will be modified to make the gravitational energy density positive.
\section*{References}
 1. M.T. Grisaru, P.van Nieuwenhuizen, and C.C. Wu, Phys. Rev. D, {\bf 12}, 397 (1975).\\ 
  2. S. Weinberg, {\sl Gravitation and Cosmology}, New York (1972).\\
 3  A. Nikishov, Part. and Nucl., {\bf 32}, 5 (2001), gr-qc/9912034.\\
 4.  L.D.Landau and E.M.Lifshitz, {\sl The classical theory of
    fields}, Addison-Wesley (1971).\\
 5. L.D.Landau and E.M.Lifshitz, {\sl Theory of Elasticity},\\
 6. A.I.Nikishov, arXiv:1011.5620; 0912.5180[gr-qc].\\
 7. L.D.Landau and E.M.Lifshitz, {\sl Fluid Mechanics}.\\
 8. J.L.Synge, {\sl Relativity: The General theory}, Amsterdam (1960).\\
 9. M.J. Duff, Phys Rev D, {\bf7}, 2317 (1973).  \\
 10. J. Schwinger, {\sl Particles, Sources and Fields},Vol.1, Addison-Wesley (1970).\\
 11. N. Rosen, Ann. Phys. (N.Y.), {\bf 63}, 127 (1970).

\end{document}